\documentclass[conference]{IEEEtran}
\IEEEoverridecommandlockouts
\usepackage{amsmath}
\usepackage{graphicx}
\usepackage{hyperref}
\usepackage{tikz}
 \usepackage{placeins} 
 \usepackage{array}
  \usepackage{caption}
  \usepackage{cite}
  \usepackage{ragged2e}
\usepackage{makecell} 
  \usepackage[caption=false,font=footnotesize]{subfig} 
\usepackage{tikz}

\usetikzlibrary{positioning}   
\usetikzlibrary{arrows.meta}   
\usetikzlibrary{fit}    

 \makeatletter
    \def\ps@IEEEtitlepagestyle{
      \def\@oddfoot{\mycopyrightnotice}
      \def\@evenfoot{}
    }
    \def\mycopyrightnotice{
      {\footnotesize 979-8-3315-7867-1/25/\$31.00~\copyright~2025 IEEE\hfill} 
     \gdef\mycopyrightnotice{}
   }

    \usepackage[mathlines,switch]{lineno}

     \makeatletter
    
    \let\old@ps@IEEEtitlepagestyle\ps@IEEEtitlepagestyle
    \def\confheader#1{%
        \def\ps@IEEEtitlepagestyle{%
            \old@ps@IEEEtitlepagestyle%
            \def\@oddhead{\strut\hfill#1\hfill\strut}%
            \def\@evenhead{\strut\hfill#1\hfill\strut}%
        }%
        \ps@headings%
    }
    \makeatother
    
    \confheader{%
            \parbox{20cm}{2025 28th International Conference on Computer and Information Technology (ICCIT)\\19-21 December, Cox’s Bazar, Bangladesh}
    }

\begin{document}

\title{\LARGE Zero-Shot to Zero-Lies: Detecting Bengali Deepfake Audio through Transfer Learning}

\author{
\IEEEauthorblockN{Most. Sharmin Sultana Samu\IEEEauthorrefmark{1}, Md. Rakibul Islam\IEEEauthorrefmark{2}, Md. Zahid Hossain\IEEEauthorrefmark{3}, \\ Md. Kamrozzaman Bhuiyan\IEEEauthorrefmark{4}, Farhad Uz Zaman\IEEEauthorrefmark{5}}
\IEEEauthorblockA{
\IEEEauthorrefmark{1}Department of CSE, BRAC University, Bangladesh.\\
\IEEEauthorrefmark{2} \IEEEauthorrefmark{3}Department of CSE, Ahsanullah University of Science and Technology, Bangladesh.\\
\IEEEauthorrefmark{4}Enosis Solutions, Bangladesh.\\
\IEEEauthorrefmark{5}Department of CSE, Southeast University, Bangladesh.
}
\IEEEauthorblockA{
Email: \IEEEauthorrefmark{1}sharminsamu130@gmail.com, \IEEEauthorrefmark{2}rakib.aust41@gmail.com, \IEEEauthorrefmark{3}zahidd16@gmail.com,\\
\IEEEauthorrefmark{4}kamrozzamaan@gmail.com, \IEEEauthorrefmark{5}farhad.zaman@seu.edu.bd
}
}

\twocolumn[
\begin{@twocolumnfalse}
\vfill

\fontsize{20}{24}\selectfont

\textbf{IEEE Copyright Notice}

\vspace{1em}

\fontsize{14}{17}\selectfont   

\noindent
\begin{minipage}{1.0\textwidth}
\justifying

\textcopyright\ 2025 IEEE. Personal use of this material is permitted. Permission from IEEE must be obtained for all other uses, in any current or future media, including reprinting/republishing this material for advertising or promotional purposes, creating new collective works, for resale or redistribution to servers or lists, or reuse of any copyrighted component of this work in other works. \\

\vspace{2em}

This work has been accepted for publication in \textbf{2025 28th International Conference on Computer and Information Technology (ICCIT)}. The final published version will be available via IEEE Xplore. \\

DOI: \textit{TBD}

\end{minipage}

\vfill
\end{@twocolumnfalse}
]

\maketitle
\begin{abstract}
The rapid growth of speech synthesis and voice conversion systems has made deepfake audio a major security concern. Bengali deepfake detection remains largely unexplored. In this work, we study automatic detection of Bengali audio deepfakes using the BanglaFake dataset. We evaluate zero-shot inference with several pretrained models. These include Wav2Vec2-XLSR-53, Whisper, PANNsCNN14, WavLM and Audio Spectrogram Transformer. Zero-shot results show limited detection ability. The best model, Wav2Vec2-XLSR-53, achieves 53.80\% accuracy, 56.60\% AUC and 46.20\% EER. We then fine-tune multiple architectures for Bengali deepfake detection. These include Wav2Vec2-Base, LCNN, LCNN-Attention, ResNet18, ViT-B16 and CNN-BiLSTM. Fine-tuned models show strong performance gains. ResNet18 achieves the highest accuracy of 79.17\%, F1 score of 79.12\%, AUC of 84.37\% and EER of 24.35\%. Experimental results confirm that fine-tuning significantly improves performance over zero-shot inference. This study provides the first systematic benchmark of Bengali deepfake audio detection. It highlights the effectiveness of fine-tuned deep learning models for this low-resource language.\\

\renewcommand
\IEEEkeywordsname{Keywords}


\end{abstract}

\section{Introduction}
Audio deepfakes create false and misleading content that can deceive individuals or influence public opinion. They are also used in cybersecurity attacks that target commercial systems. Such audio forgeries are generated with text-to-speech (TTS), voice cloning (VC) or transformation and large language models (LLMs) for text-to-audio synthesis. These technologies make synthetic speech more natural and harder to detect.

Research on audio deepfake detection aims to design effective methods with limited labeled data. Public challenges \cite{kinnunen2017asvspoof, todisco2019asvspoof, yamagishi2021asvspoof} have driven much progress and formed an active research community. Early studies focused on time-frequency features \cite{fathan2022multiresolution} or specialized neural architectures \cite{lai2019assert, aravind2020audio}. Recent works have applied large self-supervised models such as Wav2Vec2 \cite{baevski2020wav2vec} and WavLM \cite{chen2022wavlm}, showing strong results on benchmark datasets \cite{guo2024audio}. However, challenge-based methods often lack industrial robustness because they rely on small datasets and perform poorly in open conditions. Newer competitions such as the Kaggle Deep Fake Detection Challenge and AISG Trusted Media Challenge \cite{chen2022trusted} are closer to practical applications, but they emphasize video or multimodal data rather than audio alone. A further problem in this field is the open-set nature of detection. Both real and fake classes evolve, yet most studies assume a supervised setup. As new synthesis techniques emerge, a key challenge is to design systems that adapt quickly with minimal labeled data.

Most existing works in audio deepfake detection focus on English and a few high-resource languages. Studies on low-resource languages such as Bengali remain scarce. Bengali is spoken by more than 230 million people, yet resources for detecting synthetic audio in this language are limited. The lack of large annotated datasets and pretrained models tuned for Bengali makes the problem more challenging. This gap creates risks for social, financial and political misuse of deepfake audio in Bengali. To address this, we study Bengali audio deepfake detection using the publicly available BanglaFake \cite{fahad2025banglafake} dataset. Our goal is to benchmark multiple architectures under zero-shot and fine-tuned settings.

The central research question of this work is how effectively existing deep learning models can detect Bengali audio deepfakes under both zero-shot and fine-tuned settings. We ask whether large pretrained models can generalize to Bengali without task-specific training and to what extent fine-tuning improves performance when trained on a domain-specific dataset. We also investigate which architectures, including convolutional, recurrent, residual and transformer-based models, are most suitable for Bengali deepfake detection.

Our key contributions mark the following:
\begin{itemize}
    \item We present the first systematic study on Bengali audio deepfake detection using the BanglaFake dataset.
    \item We evaluate zero-shot inference with multiple pretrained models and analyze their limitations.
    \item We fine-tune six architectures, including CNN-based, ResNet, Transformer and hybrid models, for Bengali deepfake audio detection.
    \item We provide a comprehensive comparison of results using accuracy, precision, recall, F1 score, AUC and EER.
    \item We highlight the potential of fine-tuned deep learning models for building robust detection systems in low-resource languages.
\end{itemize}

The paper is organized as follows. Section II reviews existing works on audio deepfake detection and related studies in speech processing. Section III provides background study relevant to this research. Section IV describes the proposed methodology, including zero-shot inference and fine-tuning strategies with different models. Section V presents the experimental results with detailed performance analysis. Section VI concludes the paper and highlights possible directions for future research in Bengali audio deepfake detection.

\section{Related Works}
Many studies apply deep learning to audio deepfake detection. ResNet18 is widely used due to its simplicity and strong performance. It is used with feature engineering techniques such as LFCC and CQCC \cite{chen2020generalization, pan2022speaker}. Some models use attention mechanisms and recurrent layers to handle temporal dependencies. For example, a ResNet18-LSTM combination with multi-head attention improves robustness in noisy conditions \cite{pan2022speaker}. RawNet2 and its variants are applied for end-to-end detection and feature extraction \cite{muller2022does, kawa2022specrnet, kawa2022defense}. Transformer-based models like AASIST, ViT and AST show strong global context learning \cite{li2024detecting, yi2023audio, le2024continuous}. Self-supervised learning models such as Wav2Vec2, HuBERT, Whisper and AudioMAE provide good feature representations for downstream classifiers \cite{yang2024robust, guo2024audio, chen2024rawbmamba, combei2024wavlm}. Hybrid approaches using multimodal inputs also show promise. For instance, a combination of Wav2Vec2 for audio and mBERT for lyrics improves detection in musical deepfakes \cite{li2024detecting}. Some models use text-audio contrastive learning for zero-shot and multilingual scenarios \cite{ranjan2025multimodal}.

Datasets play a critical role in model evaluation and generalization. ASVspoof 2019 and 2021 are the most commonly used datasets \cite{chen2020generalization, muller2022does, muller2022human, yi2023audio, shaaban2023audio, pan2022speaker, kawa2022defense}. These datasets provide logical and physical access scenarios for controlled testing. Some studies use Fake or Real \cite{khochare2022deep, shaaban2023audio} and WaveFake \cite{kawa2022specrnet, kawa2022defense} to evaluate performance on image-based and spectrogram-based inputs. Others focus on more diverse and real-world datasets. For example, in-the-wild corpora are introduced to assess performance under uncontrolled conditions \cite{muller2022does, chen2024rawbmamba}. New datasets such as FakeMusicCaps \cite{li2024detecting} and SynHate \cite{ranjan2025synhate} expand the scope to musical and hate speech detection. SynHate includes 37 languages, supporting multilingual evaluation. Other works use large-scale paired text and audio data for low-resource languages \cite{ranjan2025multimodal}. EVDA is used to test continual learning across eight deepfake datasets \cite{chen2025region}.

Performance varies depending on model design and input features. Temporal CNN on mel-spectrograms achieves 92\% accuracy, outperforming traditional classifiers like SVM and Random Forest \cite{khochare2022deep}. Whisper-small performs well on multilingual hate speech detection, reaching 85.4\% accuracy \cite{ranjan2025synhate}. SpecRNet achieves performance close to LCNN but with fewer parameters and faster inference \cite{kawa2022specrnet}. Adaptive adversarial training reduces EER for LCNN and RawNet3 under white-box attacks \cite{kawa2022defense}. RegO improves continual learning by reducing forgetting and enhancing generalization \cite{chen2025region}. Multi-view and multi-scale feature fusion improves detection in noisy and short utterance conditions \cite{yang2024robust, chen2024rawbmamba}. Despite these advances, several models show sharp performance drops in real-world or out-of-domain scenarios \cite{muller2022does, yi2023audio, kawa2022specrnet, kawa2022defense}. Short audio clips, unseen attacks and transferability issues reduce model reliability.

Limitations are commonly reported across the reviewed papers. Many models rely on handcrafted features, which may not adapt to new spoofing methods \cite{chen2020generalization, muller2022human}. Some models are sensitive to speaker, language or recording conditions \cite{ranjan2025multimodal, ranjan2025synhate, yi2023audio}. CNN-based models often lack robustness to adversarial examples \cite{rabhi2024audio}. Training and inference are often resource-intensive, especially for large transformers or ensemble models \cite{yi2023audio, combei2024wavlm}. Generalization to novel attacks or domains remains a major challenge \cite{kawa2022defense, chen2025region}. Few studies explore interpretability or explainability of detection results \cite{shaaban2023audio}. Human performance comparisons reveal shared weaknesses between AI systems and users, especially against TTS attacks \cite{muller2022human}. Evaluation is often limited to specific datasets or attack types, reducing cross-study comparability.

Future work in this field points to several promising directions. Many studies propose stronger generalization through multi-task and continual learning \cite{zhang2024remember, le2024continuous, chen2025region}. Researchers suggest integrating diverse features, including raw audio, spectral inputs and multimodal information \cite{li2024detecting, yang2024robust, chen2024rawbmamba}. Improvements in multilingual performance are recommended, especially using self-supervised and transformer-based models \cite{ranjan2025multimodal, ranjan2025synhate, yi2023audio}. Lightweight architectures with fewer parameters are also encouraged for real-time deployment \cite{kawa2022specrnet, combei2024wavlm}. Adaptive detection mechanisms can help address evolving spoofing strategies \cite{chen2020generalization, kawa2022defense}. Advanced adversarial training and robust evaluation protocols are needed to improve security \cite{rabhi2024audio, kawa2022defense}. Standardization of datasets and benchmarks is another critical step for consistent evaluation \cite{yi2023audio}.

Furthermore, current benchmarks and datasets are limited in linguistic diversity. Notably, no prior work has explicitly focused on Bengali, a major low-resource language with millions of native speakers. This gap restricts the applicability of current systems in real-world multilingual contexts. To address these limitations, our research aims to develop a robust, Bengali-capable audio deepfake detection system with improved generalization, cross-lingual transferability and resilience to adversarial conditions.

\section{Background Study}
\subsection{Pretrained Models for Zero-Shot Audio Deepfake Detection}
Wav2Vec2-XLSR-53 \cite{conneau2020unsupervised} is a self-supervised speech representation model that learns contextual audio embeddings from raw waveform data. Whisper-small and Whisper-medium \cite{radford2023robust} are end-to-end speech recognition models that convert audio to text while producing robust audio features useful for downstream tasks. PANNsCNN14 \cite{kong2020panns} is a convolutional neural network trained on audio spectrograms for sound event detection and generates feature embeddings for classification. WavLM-Base-Plus \cite{chen2022wavlm} is a self-supervised speech model designed to capture both acoustic and linguistic information from raw audio. Audio Spectrogram Transformer (AST) \cite{gong2021ast} is a transformer-based audio spectrogram model pretrained on AudioSet, optimized to detect various sound events.
\subsection{Deep Learning Models for Fine-Tuned Bengali Audio Deepfake Detection}
Wav2Vec2-Base \cite{baevski2020wav2vec} is a self-supervised speech representation model that learns contextual audio embeddings from raw waveform and can be adapted for classification tasks. LCNN is a light convolutional neural network designed to extract time-frequency features from spectrograms for audio classification. ResNet18 is a residual network that captures hierarchical audio features through skip connections and deep convolutional layers. LCNN-Attention extends LCNN by adding an attention mechanism to focus on important regions in the audio feature maps. ViT-B16 is a transformer-based model that divides spectrograms into patches and processes them with self-attention for audio representation learning. CNN-BiLSTM combines convolutional layers for local feature extraction and bidirectional LSTM layers for capturing temporal dependencies in audio sequences.
\subsection{Evaluation Metrics}\label{A9}
Accuracy represents the percentage of correctly classified samples among all samples and indicates overall performance. Precision measures the proportion of true positive predictions among all positive predictions and reflects the reliability of positive detections. Recall calculates the proportion of true positives detected among all actual positive samples and indicates the model’s sensitivity. F1 Score is the harmonic mean of precision and recall and balances both false positives and false negatives. Equal Error Rate (EER) is the point where false acceptance rate equals false rejection rate and is a critical metric for assessing system robustness in spoof detection. A smaller EER indicates better model performance, while a larger EER shows weaker discrimination. Area Under the Curve (AUC) measures the ability of the model to distinguish between classes across different thresholds and indicates the overall discrimination capability of the classifier. A larger AUC indicates stronger discrimination capability, while a smaller AUC shows poor separability between classes.

\section{Methodology}
\subsection{Dataset and Preprocessing}
We have used the publicly available BanglaFake \cite{fahad2025banglafake} audio dataset. The BanglaFake dataset is a benchmark resource for developing and evaluating Bengali deepfake audio detection models. It contains 12,260 real and 13,260 synthetic speech samples in WAV format, each lasting approximately 6–7 seconds and recorded at 22,050 Hz. Real audio is sourced from the SUST TTS Corpus \cite{ahmad2021sust} and Mozilla Common Voice \cite{ardila2019common}, covering seven speakers, while deepfake audio is generated using a VITS-based text-to-speech model trained on the SUST TTS Corpus, designed to mimic human speech with subtle artifacts. The dataset follows the LJ Speech format which provides standardized metadata and naming conventions to ensure compatibility with existing TTS and audio processing tools. This organization supports robust training, evaluation and benchmarking of deepfake detection systems in Bengali. All audio files are resampled to 16 kHz to maintain consistency. Audio samples are truncated or zero-padded to a fixed duration depending on the model requirements. Mel-spectrograms are extracted with 64–128 Mel bands, using standard FFT and hop lengths. Spectrograms are converted to decibel scale and normalized. For image-based models, spectrograms are resized to 224×224 pixels and replicated across three channels. A custom PyTorch Dataset class handles batch loading and preprocessing.
\subsection{Zero-Shot Inference Models}
\begin{figure}[hbt!] 
    \includegraphics[width=90mm,scale=0.8]{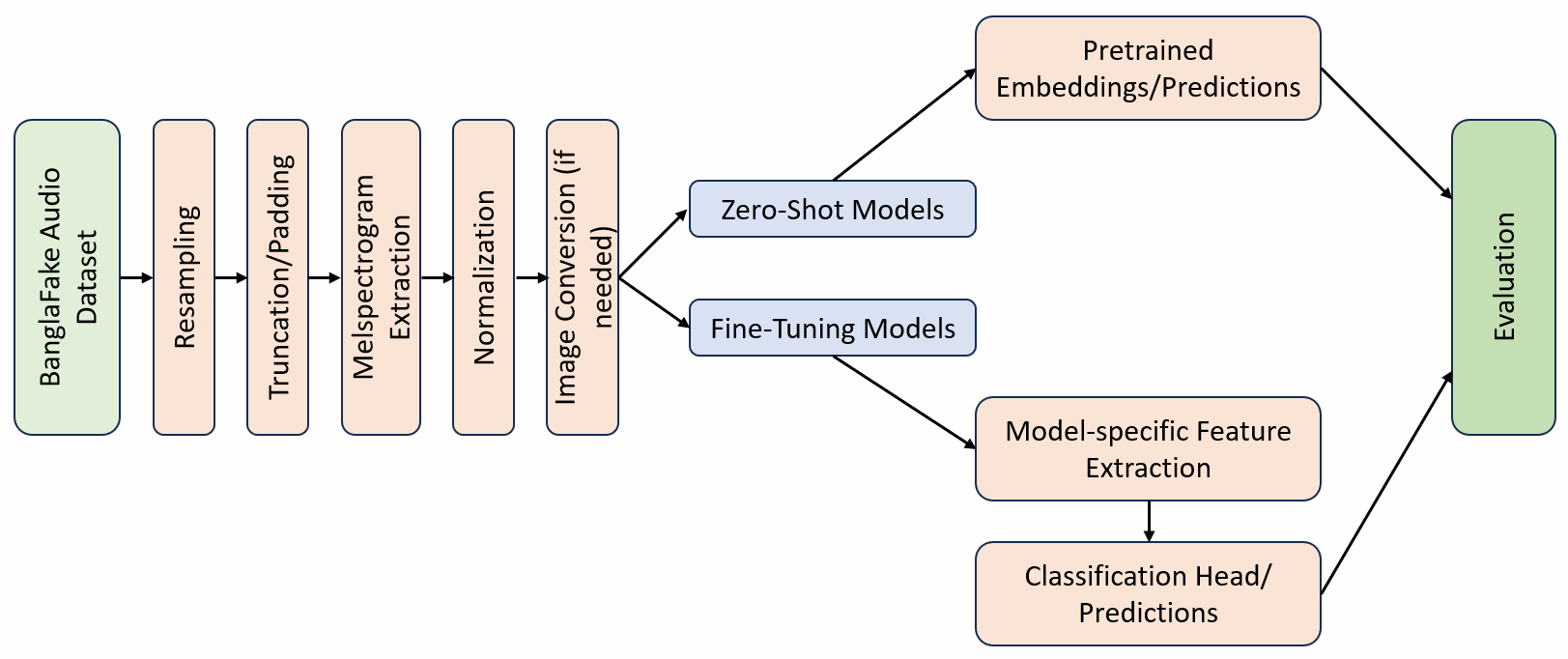}
    \caption{Proposed Methodology}
    \label{fig:gr1}
\end{figure}

We apply zero-shot inference using six pretrained models: Wav2Vec2-XLSR-53, Whisper-small, Whisper-medium, PANNsCNN14, WavLM-Base-Plus and Audio Spectrogram Transformer (AST). Each model is used without fine-tuning. Audio inputs are preprocessed as required for each model. The models generate embeddings or predictions which are evaluated directly using standard metrics. Figure \ref{fig:gr1} illustrates our proposed methodology.

\subsection{Fine-Tuning Models}
For fine-tuning, six models are trained on BanglaFake \cite{fahad2025banglafake} dataset: Wav2Vec2-Base, LCNN, LCNN-Attention, ResNet18, ViT-B16 and CNN-BiLSTM.

\begin{itemize}
    \item \textbf{Wav2Vec2-Base}: Pretrained Wav2Vec2 features are frozen. The classification head is trained on fixed-length 5-second audio samples. Outputs are binary logits for real or fake audio.
    \item \textbf{LCNN}: Mel-spectrograms are transformed into [1, 64, time] tensors. The network uses MFM layers and max pooling. A fully connected layer maps features to two classes.
    \item \textbf{LCNN-Attention}: Attention is applied over the time axis to highlight important temporal features. Outputs pass through fully connected layers for binary classification.
    \item \textbf{ResNet18}: Mel-spectrograms are expanded to 3-channel images. Features are extracted using a pretrained ResNet18 backbone. A classifier layer produces logits for deepfake detection.
    \item \textbf{ViT-B16}: Mel-spectrogram images are divided into patches for a Vision Transformer. Features are processed through the transformer encoder and the final classification head outputs binary logits.
    \item \textbf{CNN-BiLSTM}: Spectrograms are split into windows, converted to images and passed through a pretrained ResNet18 backbone. The resulting feature sequence goes through a bidirectional LSTM, followed by fully connected layers for classification.
\end{itemize}
Performance is measured using accuracy, precision, recall, F1-score, Equal Error Rate (EER) and Area Under the Curve (AUC).

\subsection{Experimental Setup}
The dataset is split into training, validation and testing sets with a ratio of 70:15:15. Two experimental settings are designed: zero-shot inference using pretrained models and fine-tuning with the BanglaFake dataset. Training setups employ Adam-based optimization, weighted or standard cross-entropy losses and early stopping criteria. Model checkpoints are preserved based on validation performance.

\begin{table}[!ht]
\centering
\caption{\label{tab:model_comparison} Comparison of deepfake audio detection models.}
\begin{tabular}{|l|c|c|c|}
\hline
\textbf{Model Name} & \textbf{\makecell{Sequence\\Modeling}} & \textbf{\makecell{Audio\\Duration}} & \textbf{\makecell{Loss\\Function}} \\
\hline
\makecell{LCNN} & No & 5 sec & Cross-Entropy \\
\makecell{LCNN-Attention} & Yes & 5 sec & Cross-Entropy \\
\makecell{ResNet18} & No & 5 sec & Binary Cross-Entropy \\
\makecell{ViT-B16} & No & 5 sec & Cross-Entropy \\
\makecell{CNN-BiLSTM} & Yes & 5 sec & \makecell{Weighted Binary\\Cross-Entropy} \\
\makecell{Wav2Vec2-Base} & Yes & 5 sec & Cross-Entropy \\
\hline
\end{tabular}
\end{table}

Table \ref{tab:model_comparison} compares several fine-tuned models used for Bengali deepfake audio detection. All models are trained with five-second audio inputs to maintain consistency. LCNN, ResNet18 and ViT-B16 rely on frame-level representations without sequence modeling, which limits their ability to capture temporal patterns in speech. LCNN-Attention, CNN-BiLSTM and Wav2Vec2-Base integrate sequence modeling, which helps them learn dependencies across time and improves detection of subtle manipulation cues. The choice of loss functions also reflects model design. Standard cross-entropy is applied in most cases as it is effective for classification tasks. ResNet18 uses binary cross-entropy, which simplifies the decision boundary for two-class detection. CNN-BiLSTM employs weighted binary cross-entropy to handle class imbalance, as fake and real audio data are often unevenly distributed.

\begin{table}[!ht]
\centering
\caption{\label{tab:training_params} Training hyperparameters for deepfake audio detection models.}
\begin{tabular}{|l|c|c|c|c|}
\hline
\textbf{Model Name} & \textbf{Optimizer} & \textbf{\makecell{Learning\\ Rate}} & \textbf{\makecell{Batch\\ Size}} & \textbf{Epochs} \\
\hline
\makecell{LCNN} & Adam & 0.0001 & 32 & 2 \\
\makecell{LCNN-Attention} & Adam & 0.0001 & 16 & 14 \\
\makecell{ResNet18} & Adam & 0.0001 & 32 & 3 \\
\makecell{ViT-B16} & Adam & 0.0001 & 8 & 3 \\
\makecell{CNN-BiLSTM} & Adam & 0.0001 & 8 & 10 \\
\makecell{Wav2Vec2-Base} & AdamW & 0.00005 & 4 & 1 \\
\hline
\end{tabular}
\end{table}

Table \ref{tab:training_params} reports the training hyperparameters used for different models in Bengali deepfake audio detection. All models use the Adam optimizer except Wav2Vec2-Base, which adopts AdamW for better regularization. The learning rate remains fixed at 0.0001 for most models to ensure stable convergence, while Wav2Vec2-Base uses a lower rate of 0.00005 due to its large parameter size. Batch sizes vary across models and reflect the trade-off between memory usage and training stability. Larger batch sizes are applied to lightweight models such as LCNN and ResNet18, while smaller batches are required for ViT-B16, CNN-BiLSTM and Wav2Vec2-Base because of higher computational demands. The number of epochs also differs and indicates model complexity and convergence behavior. LCNN and ResNet18 converge quickly within few epochs, while LCNN-Attention and CNN-BiLSTM require longer training to capture temporal patterns. Wav2Vec2-Base is trained for only one epoch due to computation resource constraint. 

\section{Result Analysis}
In this section, we present the experimental results of both zero-shot inference models and fine-tuned models.
\subsection{Zero-Shot Classification}
The zero-shot models showed moderate to low performance. The highest accuracy was achieved by Wav2Vec2-XLSR-53 at 53.8\%. PANNsCNN14 reached an accuracy of 50\% but had perfect recall (100\%), indicating it correctly identified all positive samples but also misclassified some negative samples. Whisper-small and Whisper-medium had lower accuracy (48.2\% and 46\%). Their precision and recall were equal, suggesting balanced performance but limited classification capability. WavLM-Base-Plus had zero precision and recall, indicating failure in correct positive predictions. The EER values ranged from 46.2\% to 60.0\%. The AUC values were generally low, with Wav2Vec2-XLSR-53 achieving the highest at 56.6\%. Overall, zero-shot models showed limited ability to classify correctly without task-specific training. Table \ref{tab:zeroshot_model_metrics} reports zero-shot classification results.

\begin{table}[!ht]
\centering
\caption{\label{tab:zeroshot_model_metrics} Performance comparison of pre-trained audio models for zero-shot classification. Acc, Prec, Rec, F1, EER and AUC stands for Accuracy, Precision, Recall, F1 Score, Equal Error Rate and Area Under the Curve respectively. Values are presented in percentage.}
\begin{tabular}{|l|c|c|c|c|c|c|}
\hline
\textbf{Model Name} & \textbf{Acc} & \textbf{Prec} & \textbf{Rec} & \textbf{F1} & \textbf{EER} & \textbf{AUC} \\
\hline
\makecell{Wav2Vec2-XLSR-53} & 53.8 & 52.0 & 53.8 & 52.9 & 46.2 & 56.6 \\
\makecell{Whisper-small} & 48.2 & 48.2 & 48.2 & 48.2 & 51.8 & 47.7 \\
\makecell{Whisper-medium} & 46.0 & 46.0 & 46.0 & 46.0 & 54.0 & 44.5 \\
\makecell{PANNsCNN14} & 50.0 & 50.0 & 1.0 & 66.7 & 52.6 & 43.2 \\
\makecell{WavLM-Base-Plus} & 50.0 & 0.0 & 0.0 & 0.0 & 60.0 & 36.4 \\
\makecell{AST} & 40.1 & 38.3 & 40.1 & 39.2 & 59.9 & 36.9 \\
\hline
\end{tabular}
\end{table}

\subsection{Fine-Tuned Classification}
Fine-tuned models demonstrated higher performance across all metrics. Table \ref{tab:model_metrics} presents classification results of fine-tuned models.

\begin{table}[!ht]
\centering
\caption{\label{tab:model_metrics} Performance comparison of deepfake audio detection fine-tuned models.}
\begin{tabular}{|l|c|c|c|c|c|c|}
\hline
\textbf{Model Name} & \textbf{Acc} & \textbf{Prec} & \textbf{Rec} & \textbf{F1} & \textbf{EER} & \textbf{AUC} \\
\hline
\makecell{Wav2Vec2-Base} & 65.28 & 53.37 & 98.49 & 69.23 & 30.58 & 76.17 \\
\makecell{LCNN} & 48.36 & 46.73 & 52.53 & 49.46 & 61.23 & 50.48 \\
\makecell{ResNet18} & 79.17 & 65.66 & 99.53 & 79.12 & 24.35 & 84.37 \\
\makecell{LCNN-Attention} & 78.43 & 64.94 & 99.11 & 78.47 & 23.01 & 88.48 \\
\makecell{ViT-B16} & 78.65 & 65.41 & 97.97 & 78.45 & 22.26 & 86.63 \\
\makecell{CNN-BiLSTM} & 78.49 & 64.92 & 99.53 & 78.58 & 29.76 & 79.63 \\
\hline
\end{tabular}
\end{table}

ResNet18 achieved the highest accuracy (79.17\%) and a very high recall (99.53\%). LCNN-Attention and ViT-B16 also performed well, with accuracy above 78\% and AUC above 86\%. Wav2Vec2-Base showed moderate performance with 65.28\% accuracy and high recall (98.49\%). LCNN had lower accuracy (48.36\%) but moderate F1 score (49.46\%). The EER values for fine-tuned models were much lower than zero-shot models, with ViT-B16 at 22.26\% and LCNN-Attention at 23.01\%, indicating better separation between classes. Overall, fine-tuning significantly improved classification performance and reliability compared to zero-shot approaches.

\begin{figure}[hbt!] 
    \includegraphics[width=90mm,scale=0.85]{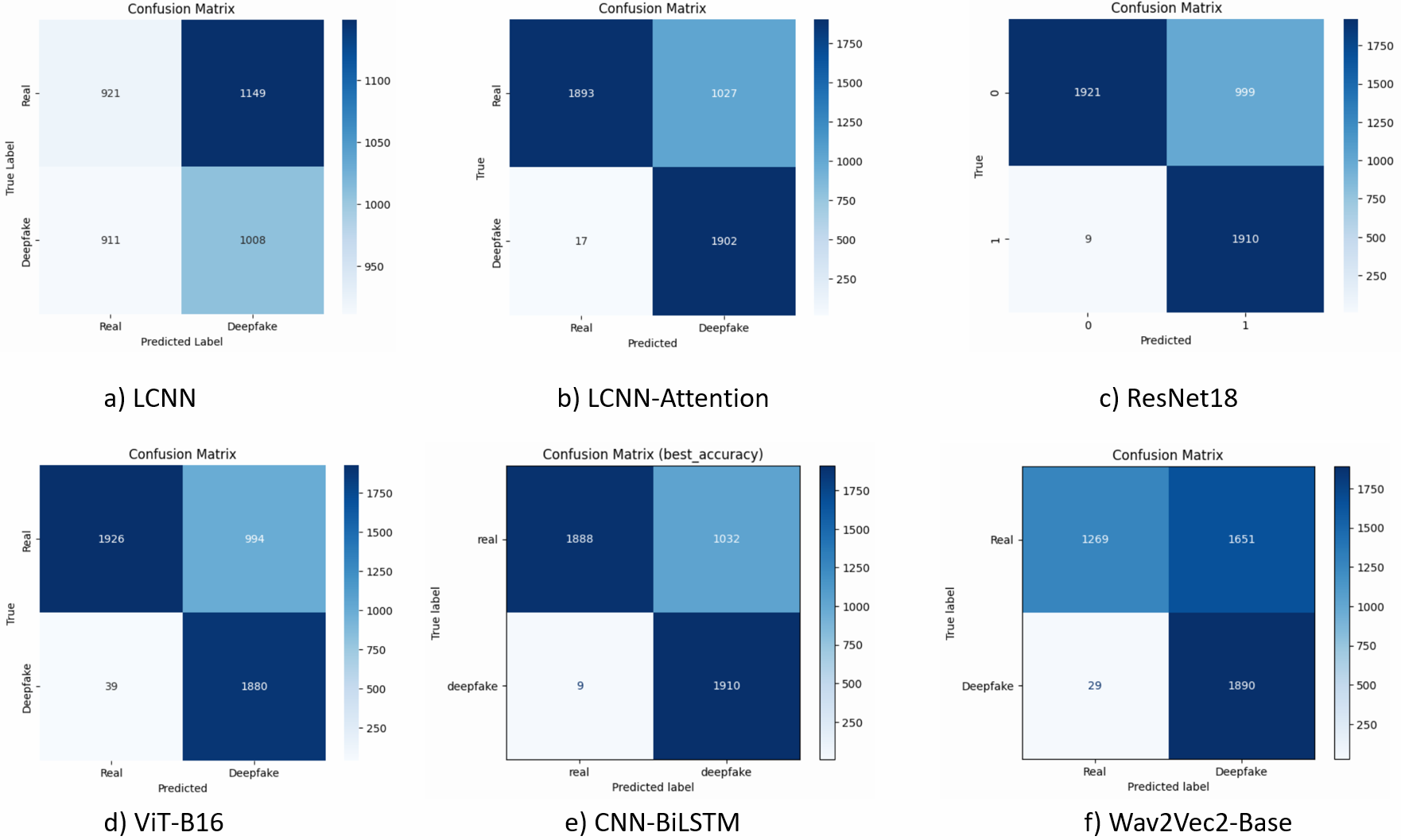}
    \caption{Confusion matrix of the fine-tuned classification models.}
    \label{fig:gr2}
\end{figure}

Figure \ref{fig:gr2} presents the confusion matrix of six fine-tuned models. LCNN shows high misclassification with balanced errors in both classes. LCNN-Attention improves detection with very few errors for deepfake but higher confusion for real. ResNet18 achieves very strong performance with minimal misclassification in both classes. ViT-B16 also performs well with slightly higher misclassification for real but low error for deepfake. CNN-BiLSTM gives results close to ResNet18 with almost no misclassification for deepfake but higher confusion for real. Wav2Vec2-Base performs poorly for real detection with high misclassification but detects deepfake with strong accuracy.

\begin{figure}[hbt!] 
    \includegraphics[width=90mm,scale=0.85]{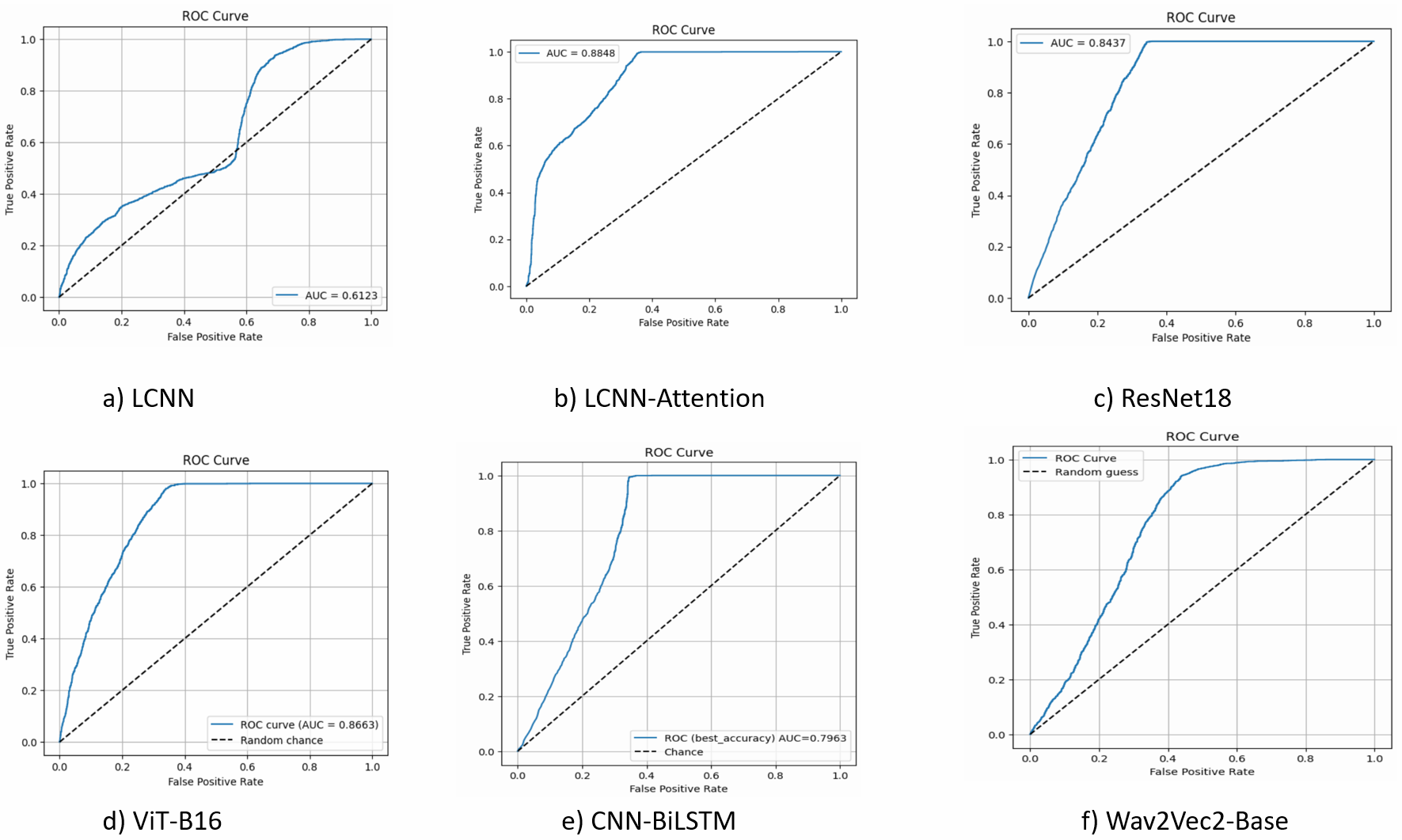}
    \caption{ROC curve of the fine-tuned classification models.}
    \label{fig:gr3}
\end{figure}

Figure \ref{fig:gr3} presents the ROC curves of six fine-tuned models. LCNN shows poor separability with an AUC of 50.48\% and the curve remains close to the diagonal. LCNN-Attention performs strongly with an AUC of 88.48\% and the curve rises steeply toward the top left corner. ResNet18 achieves an AUC of 84.37\% with a smooth curve showing consistent discrimination ability. ViT-B16 provides one of the best results with an AUC of 86.63\% and the curve remains close to the ideal boundary. CNN-BiLSTM produces an AUC of 79.63\% with moderate classification strength and less steep rise. Wav2Vec2-Base shows weaker performance as the curve lies nearer to the diagonal with limited separation despite partial rise.

\begin{figure}[hbt!] 
    \includegraphics[width=90mm,scale=0.85]{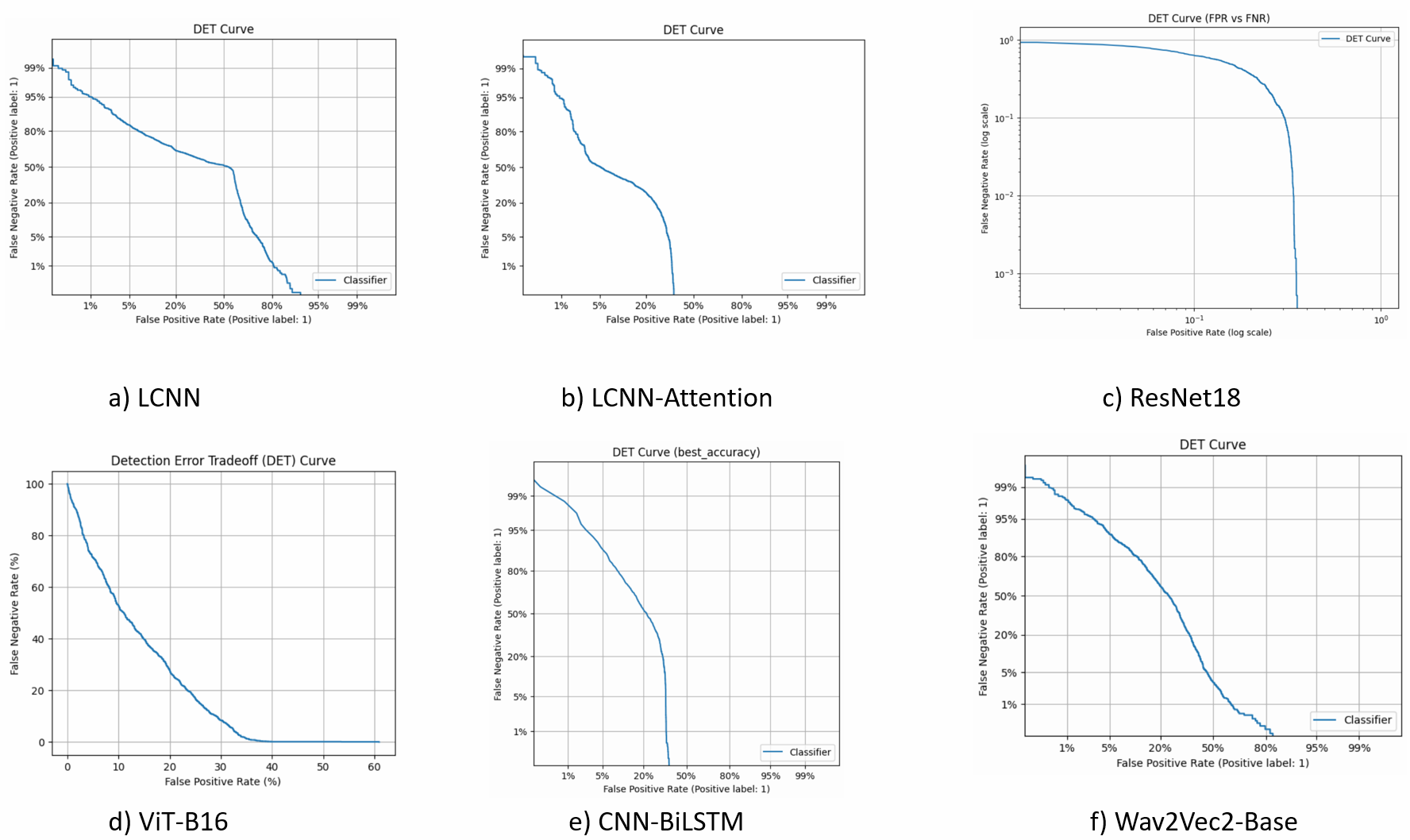}
    \caption{DET curve of the fine-tuned classification models.}
    \label{fig:gr4}
\end{figure}

The Detection Error Tradeoff (DET) curves in Figure \ref{fig:gr4} show the error tradeoff across six models. LCNN presents high false negative rates across most thresholds, indicating weak separability. LCNN-Attention reduces errors with a smoother curve that drops faster, showing better balance between false positives and false negatives. ResNet18 achieves the strongest performance with a steep curve in logarithmic scale and very low error rates at optimal thresholds. ViT-B16 maintains a stable curve with gradual decline, reflecting reliable classification and low error at moderate thresholds. CNN-BiLSTM shows improved detection compared to LCNN, with a sharp drop in errors at lower false positive rates, but less steep than ResNet18. Wav2Vec2-Base produces higher error rates with a slower decline, indicating weaker discrimination power compared to vision-based models.

\section{Conclusion and Future Works}
We present the first systematic benchmark for Bengali deepfake audio detection in this study. The results show that zero-shot inference with pretrained models provides limited effectiveness. Fine-tuned models achieve significant improvements across all metrics. ResNet18 gives the best performance among the tested architectures. The findings confirm that fine-tuning is necessary for robust detection in a low-resource setting. The study highlights the challenges of detecting deepfakes in Bengali audio and demonstrates the potential of deep learning methods for addressing this problem. Future work should expand the dataset to cover more speakers and diverse synthesis techniques. Cross-lingual transfer learning can be explored to improve performance in low-resource conditions. Robustness against adversarial attacks and unseen deepfake generation methods should be investigated. Lightweight models are required for real-time applications and deployment in resource-constrained environments. Future research should also integrate prosodic and linguistic features with acoustic cues to enhance detection reliability.

\bibliographystyle{ieeetr} 

\addcontentsline{toc}{chapter}{References}
\bibliography{ref}

\end{document}